\documentclass[prb,preprint,aps,showpacs,preprintnumbers,amsmath,amssymb]{revtex4}

\usepackage{graphicx}
\usepackage{bm}
\usepackage{color}

\begin{document}

\title{Nanosized superparamagnetic precipitates in cobalt-doped ZnO}

\author{Matthias Opel}
     \email{Matthias.Opel@wmi.badw.de}
\author{Karl-Wilhelm Nielsen}
\author{Sebastian Bauer}
\author{Sebastian T. B. Goennenwein}
\author{Rudolf Gross}
    \email{Rudolf.Gross@wmi.badw.de}

\affiliation{Walther-Mei{\ss}ner-Institut, Bayerische Akademie der Wissenschaften,
             85748 Garching, Germany}

\author{J\'{u}lio C. Cezar}
\affiliation{European Synchrotron Radiation Facility, 38043 Grenoble Cedex 9, France}%

\author{Dieter Schmeisser}
\affiliation{Angewandte Physik II, Brandenburgische Technische Universit\"{a}t Cottbus, 03046 Cottbus, Germany}%

\author{J\"{u}rgen Simon}
\author{Werner Mader}
\affiliation{Institut f\"{u}r Anorganische Chemie, Rheinische Friedrich-Wilhelms-Universit\"{a}t
             Bonn, 53117 Bonn, Germany}

\date{\today}

\begin{abstract}  
The existence of semiconductors exhibiting long-range ferromagnetic ordering at
room temperature still is controversial. One particularly important issue
is the presence of secondary magnetic phases such as clusters, segregations,
etc... These are often tedious to detect, leading to contradictory
interpretations. We show that in our cobalt doped ZnO films grown
homoepitaxially on single crystalline ZnO substrates the magnetism unambiguously
stems from metallic cobalt nano-inclusions. The magnetic behavior was
investigated by SQUID magnetometry, x-ray magnetic circular dichroism, and AC
susceptibility measurements. The results were correlated to a detailed
microstructural analysis based on high resolution x-ray diffraction, transmission
electron microscopy, and electron-spectroscopic imaging.
No evidence for carrier mediated ferromagnetic exchange between
diluted cobalt moments was found. In contrast,
the combined data provide clear evidence that the observed room
temperature ferromagnetic-like behavior originates from nanometer sized
superparamagnetic metallic cobalt precipitates.
\end{abstract}


\maketitle

\section{Introduction}

Materials combining different functionalities have attracted huge interest due
to the ever growing demands in applications. Regarding spintronics
\cite{Wolf:2001a,Zutic2004}, semiconductors that are ferromagnetic above room
temperature (RT) are in the focus of current research, as they combine
ferromagnetic exchange with versatile electronic properties. They certainly
would pave the way to novel devices with new functionalities. Dilute magnetic
semiconductors (DMS) in which itinerant charge carriers mediate a ferromagnetic
coupling between the moments of diluted transition metal (TM) ions, have been
considered as promising materials \cite{Dietl2000}. While ferromagnetism has
first been observed in Mn-doped InAs \cite{Munekata1989} and GaAs
\cite{Ohno1996} the Curie temperatures $T_{\rm C}$ in those materials are still
well below RT. In contrast, RT ferromagnetism has been predicted for wide
bandgap TM-doped semiconductors such as ZnO:TM or GaN:TM \cite{Dietl2000}.
Although these materials have been studied extensively during the last years
their magnetic properties are still under debate. For ZnO:TM, hole-mediated
ferromagnetism was originally suggested \cite{Dietl2000} whereas in the
majority of experimental studies RT ferromagnetism has been found in $n$-type
material
\cite{Venkatesan2004,Rode2003,Chakraborti2007,Liu2007,Pan2007,Nielsen2006,Chambers:2006a,Kittilstved:2006b}.
In turn, recent theoretical models explain ferromagnetic coupling in
terms of bound magnetic polarons \cite{Coey2005,Kittilstved:2006c} or a ligand-to-metal charge
transfer \cite{Liu2005,Kittilstved:2006a}. While a large number of experimental
studies seem to provide evidence for carrier-mediated RT ferromagnetism in ZnO:TM, also
the absence of any ferromagnetic coupling has been reported
\cite{Kolesnik2004,Pacuski2006,Yin2006,Sati2007,Chang2007}, thereby further
fueling the controversy about the occurrence and nature of ferromagnetism in
ZnO:TM. Furthermore, calculations show a magnetic moment at oxygen rich
surfaces in e.g. ZrO$\mathrm{_2}$ or Al$\mathrm{_2}$O$\mathrm{_3}$
\cite{Gallego2005} which even calls into question the necessity of TM-doping
for ferromagnetism. Finally, unexpected magnetic coupling has also been found
in nominally undoped oxides like HfO$_2$ or TiO$_2$
\cite{Coey2004,Hong2006}.

Another important issue concerning DMS is the formation of ferromagnetic
clusters. While they have long been regarded as detrimental for spintronic
applications it has recently been shown that they may be utilized to control
high temperature ferromagnetism in semiconductors and to tailor spintronic
functionalities \cite{Kuroda:2007a}. TM-doped ZnO is known to form nanosized
(inter)metallic inclusions \cite{Wi2004,Zhou2006,Li2007,Sudakar2007} which may
be responsible for the observed RT magnetic response. The same is true for
other TM-doped semiconductors like Mn-doped Ge \cite{Ahlers2006,Jaeger2006}.
Therefore, an unambiguous clarification of the origin of magnetism requires a
systematic combined magnetic and microstructural analysis. We have performed
such a study on cobalt-doped ZnO. For magnetic characterization, we used SQUID
magnetometry as well as x-ray magnetic circular dichroism (XMCD) and AC
susceptibility measurements. The advantage of XMCD is the independent and
element-specific determination of spin and orbital magnetic moments on an
atomic level. In contrast to previous studies, we performed XMCD in both the
total electron and fluorescence yield modes allowing to distinguish between
surface and bulk magnetic properties. Our data provide clear evidence that the
observed room temperature magnetism is not related to a bulk homogeneous DMS,
but can rather be explained by the presence of superparamagnetic metallic
cobalt precipitates. This is further confirmed by careful x-ray diffraction and
high-resolution transmission electron microscopy, making the nanosized cobalt
inclusions visible.

\section{Sample Fabrication}

The epitaxial Zn$_{0.95}$Co$_{0.05}$O thin films studied here
were grown by pulsed laser deposition from a stoichiometric
polycrystalline target using a KrF excimer laser (248~nm) at a repetition rate
of 2~Hz with an energy density at the target of 2~J/cm$^2$. The thin films were
deposited on single crystalline, $c$-axis oriented ZnO(0001) substrates at
temperatures $T_{\rm G}$ between
$300^\circ$C and $600^\circ$C in pure argon atmosphere at a pressure of
$4 \times 10^{-3}$~mbar. During deposition, the film growth was monitored by
in-situ RHEED \cite{Gross:2000a}.
The structural analysis of the samples was performed in a
Bruker-AXS four circle diffractometer (D8 Discover) using Cu K$_{\alpha 1}$
x-ray radiation. High-resolution x-ray diffraction ($\omega$-2$\theta$)
scans in out-of-plane direction and reciprocal space
mappings of the $(10\bar{1}1)$ reflection reveal an excellent crystalline
quality of the films. The $c$-axis lattice parameters were found to range
between 5.22 and 5.32\,\AA. The mosaic spread indicated by the full width at
half maximum of the rocking curves of the (0002) reflection was as low as
$0.02^{\circ}$. More details are given elsewhere \cite{Nielsen2006}.
As the samples were grown in the absence of oxygen they are highly conductive.
Their room temperature resistivity is in the order of $1\,{\rm \Omega cm}$.
Although similar data have been found for all films grown
at different temperatures, in the following we will focus on Zn$_{0.95}$Co$_{0.05}$O
thin films grown at $T_{\rm G} = 400^\circ$C and
$500^\circ$C with a thickness of 350~nm, for clarity.

\section{Magnetic Characterization}

Magnetization and AC susceptibility were measured in a Quantum
Design superconducting quantum interference device (SQUID) magnetometer (MPMS
XL-7) with a magnetic field of up to 7~T applied in plane.
The magnetization $M$ as a function of the magnetic field $H$ shows an
``S''-shaped behavior at room temperature, as shown in Fig.1(a). The
data have been corrected for the linear diamagnetic contribution of the
substrate. $M$ saturates at $\mu_0H \simeq 3$~T showing values of $M_{\rm S} =
1.02$ and 1.95 Bohr magnetons ($\mu_{\rm B}$) per Co atom for the samples grown
at $400^\circ$C (green) or $500^\circ$C (blue), respectively. We note that the
shape of the $M(H)$ curves is about the same for both deposition temperatures.
Similar RT magnetization data for doped ZnO thin films have been reported in
literature
\cite{Venkatesan2004,Chakraborti2007,Liu2007,Pan2007,Zhou2006,Sudakar2007}. It
is tempting to interpret these curves as evidence for RT ferromagnetism as they
cannot be explained by simple paramagnetic Brillouin functions for Co$^{2+}$ in
the high-spin ($S = 3/2$) or low-spin state ($S = 1/2$) due to their large
slopes at zero field. However, within experimental error our data do not show
any remanent RT magnetization at zero field (see inset in Fig.1(a)).
This observation is consistent with literature
\cite{Venkatesan2004,Sudakar2007}. This lack of any observable magnetic
hysteresis makes an interpretation in terms of a dilute \emph{ferromagnetic}
semiconductor questionable.

As shown in Fig.1(a), it is easily possible to fit the data by a
Langevin function
\begin{equation}
    M(B) = M_{\rm S} \left(\coth \frac{\mu B}{k_{\rm B} T} - \frac{k_{\rm B} T}{\mu B} \right)
    \label{eq:langevin}
\end{equation}
with the magnetic induction $B$, the Boltzmann constant $k_{\rm B}$, the
measuring temperature $T = 300$~K, and the moment $\mu$ of (super-)paramagnetic
particles within the thin film. Fitting the data (solid lines in
Fig.1(a)) gives $\mu = 2370\,\mu_{\rm B}$ and $5910\,\mu_{\rm B}$ for
the films grown at $400^\circ$C and $500^\circ$C, respectively. The good fits
suggest that the measured magnetization curves can be consistently explained by
the presence of superparamagnetic particles in the ZnO matrix with average
magnetic moments of $2370\,\mu_{\rm B}$ and $5910\,\mu_{\rm B}$, as already
suggested earlier \cite{Wi2004}. This calls into question the widely accepted
interpretation of the RT magnetization data of cobalt-doped ZnO thin film
samples. In the vast majority of publications, magnetization curves similar to
those shown in Fig.1(a) have been regarded as proof for the existence
of carrier mediated RT ferromagnetic coupling between dilute Co$^{2+}$ moments
in the ZnO matrix. However, the perfect fit of the data by a Langevin function
and the very small or even absent remanent magnetization shows that an
alternative interpretation of the magnetization curves in terms of nanometer
sized superparamagnetic particles with average moments of a few $1000\,\mu_B$
may be more adequate. We note that even a finite remanence would be consistent
with this interpretation since clusters with larger diameter may be blocked
already at RT. With the saturation magnetization of $1.7\,\mu_{\rm B}/{\rm Co}$
for metallic Co at room temperature \cite{OHandley2000} and assuming a
hexagonal crystallographic structure, the diameter of metallic Co clusters in
our samples is determined to about 3~nm ($T_G=400^\circ$C) and 4~nm
($500^\circ$C) to yield the moments given above.

To further clarify the nature of magnetism in our cobalt-doped ZnO films we
have performed zero field- (ZFC) and field-cooled (FC) measurements of the
temperature dependence of the magnetization. The results are presented in
Fig.1(b), where we have plotted $M(T)$ for the samples cooled down
from room temperature to 4~K at 0~T (ZFC) or 7~T (FC), respectively. Then,
$M(T)$ was measured while warming up the sample at a small measuring field of
10~mT. For both samples, there is a clear difference between the ZFC and the FC
data at low temperatures. In particular, the ZFC curves show pronounced maxima
at around 15~K and 38~K for the samples grown at $400^\circ$C or $500^\circ$C,
respectively. These maxima which are absent in the FC measurements might
originate from domain formation in a ferromagnetic ZnO:Co thin film. However,
more likely they can be explained in terms of the blocking of superparamagnetic
metallic Co nanoparticles within a diamagnetic ZnO matrix \cite{Neel1953}.
This explanation is confirmed by plotting the data from zero field-cooling as a
function of the inverse temperature up to 375~K (see inset of Fig.1(b)).
Well above the blocking temperature, the magnetization straightly follows
the Curie law ($M \propto T^{-1}$) valid for pure paramagnets.

Additional information can be obtained from the temperature dependence of the
real part of the AC susceptibility (Fig.2). This quantity was measured
on warming up the sample after cooling down in zero magnetic field. The small
AC magnetic field with amplitude $\mu_0H_{\rm AC}=0.5$~mT and frequency $f =
0.1, 1,$ and $10$~Hz was applied parallel to the film plane. The $\chi_{\rm
AC}^\prime(T)$ curves show pronounced maxima at about the same temperatures
where the FC and ZFC $M(T)$ curves start to deviate from each other (cf.
Fig.1). The positions of the maxima shift to higher temperatures with
increasing driving frequency. This is expected if the measured $\chi_{\rm
AC}^\prime$ signal originates from superparamagnetic particles
\cite{Dormann1997}. A quantitative measure of the frequency shift is given by
the relative shift of the peak temperature per decade shift in frequency,
$\Delta T_{\rm B}/T_{\rm B} \Delta \log_{10}f$. For the samples grown at
$400^\circ$C and $500^\circ$C, we obtain values of 0.06 and 0.10, respectively.
According to Dormann \textit{et al.} \cite{Dormann1997,Dormann1999}, these
values point to the presence of magnetic particles in between the
non-interacting and the weakly interacting regime. Additionally, the frequency
dependence of $T_{\rm B}$ for our thin films can be well described using the
N\'{e}el-Arrhenius law
\begin{equation}
    f = f_0 \exp\left(-\frac{E_{\rm a}}{k_{\rm B}T}\right)
    \label{eq:Neel-Arrhenius}
\end{equation}
valid for superparamagnetic particles \cite{Dormann1997,Dormann1999}, with an
activation energy $E_{\rm a}$ and a characteristic frequency $f_0$. Fitting the
data (see insets of Fig.2), we obtained $E_{\rm a}/k_{\rm B} = 580$~K
and 910\,K for the samples grown at $400^\circ$C and $500^\circ$C,
respectively. The derived blocking temperatures and activation energies agree
well with those expected for metallic Co nanoparticles with a diameter of
3-4\,nm \cite{OHandley2000,Luis:2002a}. Evidently, both DC and AC magnetization
measurements point to the existence of superparamagnetic particles within our
Zn$_{0.95}$Co$_{0.05}$O thin films. Unfortunately, in the vast majority of
literature neither the FC and ZFC magnetization curves nor the AC
susceptibility data are shown so that it is difficult to rule out superparamagnetism.

A powerful tool for clarifying the microscopic origin of ferromagnetism is
x-ray magnetic circular dichroism (XMCD) spectroscopy. This element specific
technique allows to obtain microscopic information on the magnetic ordering of
the Co magnetic moments alone. We performed x-ray absorption near edge
spectroscopy (XANES) and XMCD at the European synchrotron radiation facility
(ESRF) in Grenoble (France), beamline ID08, at temperatures of 10~K and 300~K.
Magnetic fields of up to 4~T were applied parallel or anti-parallel to the
incident x-ray beam. Two APPLE II undulators provide a left (lcp) or right
circular polarization (rcp) of almost 100\% for the incoming photons. The
samples were aligned at an angle of $75^\circ$ between the surface normal and
the incident light. Both the fluorescence yield (FY) and the total electron
yield (TEY) signals were detected simultaneously in a photon energy range from
765 to 815\,eV. The energy resolution of the spherical grating monochromator is
$\Delta E / E = 5 \times 10^{-4}$ at 850~eV. The XMCD spectra were obtained as
direct difference between consecutive XANES scans at the Co $L_3$ and $L_2$
edges recorded with opposite helicities of the x-rays. Each measurement
consists of eight XANES scans taken at constant magnetic field with four
spectra taken for left (lcp) and four for right circularly polarized (rcp)
light, respectively. As the escape depth for the secondary electrons is much
shorter \cite{Naftel1999} than for the fluorescence photons \cite{Henke1982},
the TEY mode probes the surface of the sample while the FY mode is more sensitive
to the bulk.

The XANES scans were taken at
constant magnetic field for left (lcp) and right circularly polarized (rcp)
light, respectively. The corresponding intensities are denoted by $I_+$ (lcp)
and $I_-$ (rcp). In the following, we will focus on thin films grown at
$400^\circ$C. The data were evaluated in the following four steps:
First, the XANES intensity at energies below the Co $L_3$ edge (772\,eV)
was set to zero by subtracting a constant background. Second, at 810\,eV,
i.e. above the $L_2$ edge, the XANES intensity was normalized to unity.
This is motivated by the assumption of no Co induced absorption below the
$L_3$ edge and a constant non-resonant absorption above the $L_2$ edge. In
this way, the spectra were corrected for time-dependent drifts in the measurement
setup. Averaging all corrected spectra results in the quantity $(I_++I_-)/2$
shown in Figs.3(a) and (b) for FY and TEY, respectively. Third, in order to
remove the non-resonant background, step functions sketched as dashed lines in
Figs.3(a) and (b) were subtracted. The positions of the steps were set to the center positions
of the $L_3$ and $L_2$ edges, respectively, with a fixed height ratio of $2:1$
according to the number of states available for non-resonant absorption \cite{Chen1995}.
Finally, the XMCD signal was determined by subtracting the corrected rcp from the lcp XANES
spectra. In Figs.3(c) and (d) the averaged difference $(I_+-I_-)/2$ is
shown for FY and TEY, respectively.
We note that the TEY XMCD signal at the $L_2$ edge is very weak
and may become smaller than the experimental error. This is consistent with
literature, where calculations for Co$^{2+}$ in the high spin state show only a very weak
$L_2$ signal \cite{Mamiya2006}. Unfortunately, this may result in the fact
that there is no longer any observable sign change in the XMCD signal at the $L_2$ edge
(Fig.3(d)).
We also note that some of the step functions used for background subtraction go above
the experimental XANES data (see Figs.3(a) and (b)) what might be considered
unphysical. We also used step functions staying below the experimental data in
the relevant energy range. However, this results in the problem that the
background curve does not meet the data curve above 810\,eV. A possible remedy
would be the subtraction of an additional linear background which is however
also difficult to justify. Since in any way the use of different background
functions only results in a variation of about 30\% in the derived magnetic
moments, but not in the magnetic field dependence of the moments, we used the
step functions shown in Figs.3(a) and (b).

Comparing the XMCD spectra for FY and TEY
shows significant differences. Figs.3(e) and (f) display the region
around the Co $L_3$ edge on an enlarged scale at room (e) and at low
temperature (f), respectively. The TEY spectra exhibit a pronounced fine
structure: five peaks are visible at photon energies of 779.0, 779.6, 780.2,
781.6, and 782.0\,eV. The measured data correspond well with calculations for
Co$^{2+}$ (shown in black) taken from Kobayashi \textit{et al.}
\cite{Kobayashi2005}, in particular at low temperature. Recalling the fact that
the typical escape depth of the secondary electrons in TEY mode is less than
10~nm \cite{Naftel1999}, this correspondence indicates the existence of
Co$^{2+}$ ions near the surface of the Zn$_{0.95}$Co$_{0.05}$O film. In
contrast, there is only a very weak fine structure in the FY spectra which
completely disappears at room temperature. The FY spectra are close to XMCD
data published for metallic cobalt (shown in green) by Mamiya \textit{et al.}
\cite{Mamiya2006}. The correspondence between the data sets is striking. As the
fluorescence photons have a larger escape depth of approximately 100~nm
\cite{Henke1982}, this observation provides further clear evidence for the
presence of metallic Co precipitates in the ``bulk'' of Zn$_{0.95}$Co$_{0.05}$O
thin films.

Applying the magnetooptical sum rules \cite{Thole1992,Carra1993}, we have derived
the effective spin moment $m_\mathrm{s,eff}$ and orbital moment $m_\mathrm{l}$ of
Co from the XMCD spectra for different magnetic fields following Chen's
approach \cite{Chen1995} and using $n_{\rm 3d}=7$ for the 3d electron occupation number.
While the decay of the excited state via secondary electrons varies by only about 20\%
for the $L_3$ and $L_2$ edges the fluorescence decay may differ up to 400\%
due to self-absorption effects \cite{deGroot1994}. Therefore,
the FY-XANES spectra were corrected prior to the application of the sum rules by
scaling the $L_3$ signal. The scaling factor is determined at fields below 1\,T
and is given by the area ratio of
$\frac{\rm XANES_{L_3}^{TEY} / XANES_{L_2}^{TEY}}{\rm XANES_{L_3}^{FY} / XANES_{L_2}^{FY}}$.
The results are shown in Fig.4(a) for the FY mode at room temperature. The
``S''-shaped field dependence of the derived effective spin magnetic moment
$m^\mathrm{FY}_\mathrm{s,eff}(H)$ clearly reminds of the $M(H)$ curves shown in
Fig.1. In fact, the $m^\mathrm{FY}_\mathrm{s,eff}(H)$ and the $M(H)$
curves from the sample grown at $400^\circ$C match well except for a scaling
factor of 2.5. The corresponding saturation magnetization derived from the
FY-XMCD signal is only 0.4\,$\mathrm{\mu_B/Co}$. This can be explained assuming
the presence of a ``dead magnetic layer'' of some ten nanometers at the surface
where the metallic cobalt clusters are oxidized. We suggest that this layer
only marginally affects the overall SQUID signal of the 350\,nm thick sample
but may significantly reduce the magnetization derived from the FY-XMCD signal
because of the FY-XMCD probing depth of only 100~nm \cite{Henke1982}. In
Fig.4(a), also the orbital magnetic moment
$m^\mathrm{FY}_\mathrm{l}(H)$ is shown. It is aligned parallel to
$m^\mathrm{FY}_\mathrm{s,eff}(H)$ with values up to about
0.07\,$\mathrm{\mu_B/Co}$. It is obvious that $m^\mathrm{FY}_\mathrm{l}(H)$
cannot account for the discrepancy to the saturation moment measured by SQUID
magnetometry.

In the following, we will further discuss the effective spin magnetic moment.
In Fig.4(b), we compare the $m_\mathrm{s,eff}(H)$ values obtained in
the FY and TEY mode. While the FY signal nicely follows the overall
magnetization, for the TEY mode only very small $m^\mathrm{TEY}_\mathrm{s,eff}$
values are obtained. They are comparable to the average paramagnetic moment
expected for Co$^{2+}$ ions in the high-spin state ($S = 3/2$). We would like
to point out that in general the quantitative analysis of the TEY signal is
difficult and involves considerable errors due to the small signal level.
However, as shown in the inset of Fig.4(b),
$m^\mathrm{TEY}_\mathrm{s,eff}(H)$ follows roughly the Brillouin function
$B_{3/2}$ for $g = 2$, $S = 3/2$ and $T=300$\,K, expected for isolated
Co$^{2+}$ ions. Together with the observed fine structure of the TEY-XMCD
signal from Fig.3(e,f), this result confirms the presence of
magnetically uncoupled Co$^{2+}$ ions within the small escape depth of the
secondary electrons, i.e. the surface layer of the samples. This further
supports the notion that metallic Co clusters at the sample surface are
oxidized. Similar results are reported from TEY-XAS measurements
\cite{Wi2004,Liu2007}. However, our results obtained in the FY mode clearly
point to the presence of metallic Co inclusions within the (larger) escape
depth of the fluorescence photons, i.e. the ``bulk'' of the
Zn$_{0.95}$Co$_{0.05}$O thin films.

\section{Structural Analysis}

Since the magnetic characterization of the Zn$_{0.95}$Co$_{0.05}$O thin films
provided clear evidence for the presence of metallic Co clusters we performed a
detailed microstructural analysis to directly confirm this fact. Fig.5
shows x-ray diffraction diagrams from Zn$_{0.95}$Co$_{0.05}$O thin films grown
at $400^\circ$C and $500^\circ$C. At first glance, the $\omega$-$2\theta$ scans
look impeccable, as expected for phase-pure Zn$_{0.95}$Co$_{0.05}$O thin films
with high crystalline quality. However, as shown in the inset, an additional
weak reflection in the $\omega$-$2\theta$ scan appears around $2 \theta =
44.2^\circ$ after strongly increasing the integration time up to 400\,s per
point. The position of this peak agrees best with what is expected for the
(111) reflection of fcc or the (0002) reflection of hcp cobalt. Furthermore,
the ZnCo$_2$O$_4$ (400) and the Co$_3$O$_4$ (400) peaks are close. Assuming
that this reflection originates from metallic Co clusters as suggested by XMCD,
we can use Scherrer's expression \cite{Cullity:2001a} to derive the average
cluster size from the FWHM of the diffraction peak. Doing so, we obtain
diameters of 2.2 and 3.4\,nm for the samples grown at $400^\circ$C and
$500^\circ$C, respectively. These values agree well with the values 3 and 4\,nm
obtained earlier from the fits of the $M(H)$ curves by Langevin functions.

To complete the microstructural analysis, we have performed a detailed
transmission electron microscopy (TEM) study of the thin film samples in cross
section. TEM was carried out using a Philips
CM300UT field-emission transmission electron microscope (FEG-TEM)
equipped with an electron energy imaging filter (GIF, Gatan Inc.).
The images shown here were calculated as averages of two single
exposures, each taken with an exposure time of 80\,s.
The bright field TEM image of Zn$_{0.95}$Co$_{0.05}$O grown at
$500^\circ$C (Fig.6(a)) shows characteristic contrasts spread all over
the deposited film. The contrasts definitely do not originate from the ion
milling process used for TEM sample preparation. The comparison with pure ZnO
thin films grown under the same conditions indicates that those defects are
correlated with the incorporation of Co into the ZnO film. The regions with
contrast different from that of ZnO are on a typical scale of 5~nm (yellow
circles). Analysis yields the observed contrast to be a typical Moir\'{e} contrast
originating from overlapping crystals with different structure which can be
contributed to metallic cobalt with orientation like ZnO.

The chemical composition of those regions was evaluated using energy-filtering
TEM (EFTEM). Using the three-window method \cite{Hofer1995} at the Co-L
ionisation edge, the Co distribution map of the same region as shown in
Fig.6(a) is generated (Fig.6(b)). A significant cobalt
enrichment is observed exactly in the regions of the Moir\'{e} contrasts whereas
the Co signal in the ZnO matrix is below noise level (Fig.6(b)). Also
the shape and the size of these Co enriched regions clearly correlate with the
structural features seen in the bright field image. Our HRTEM results provide
direct evidence for the presence of Co clusters in cobalt-doped ZnO
thin films. In order to distinguish whether the clusters consist of metallic Co
or some cobalt oxide we have performed additional EFTEM studies of zinc and
oxygen. The elemental maps generated at both the Zn-L~edge and the O-K~edge
show a depletion of the corresponding elements in the regions of the Co rich
clusters. The observed decrease of the O and the Zn signals in these regions
further supports our conclusion that there are clusters consisting of metallic
Co embedded in the cobalt-doped ZnO film. We may note that the HRTEM analysis
yields a typical diameter of 5\,nm for the Co cluster size in the film grown at
$500^\circ$C. This again corroborates the values derived from both the magnetic
characterization and the x-ray diffraction diagrams.

\section{Conclusions}

From our comprehensive study of both the magnetic and structural properties of
epitaxial Zn$_{0.95}$Co$_{0.05}$O thin films we can draw several important
conclusions regarding the nature of magnetism in these films. First, the
macroscopic magnetization obtained by SQUID magnetometry, the temperature
dependence of the magnetization in the FC and ZFC mode, as well as the AC
susceptibility can be consistently explained by superparamagnetic particles
with magnetic moments of several $1000\,\mu_B$. Second, the XMCD spectra
obtained in TEY mode show a multiplet structure pointing to the existence of
Co$^{2+}$ ions in the surface layer of the samples. However, in FY mode probing
the bulk of the films the spectra are smooth, resembling those of metallic
cobalt. This indicates the presence of metallic cobalt inclusions in the
Zn$_{0.95}$Co$_{0.05}$O thin films which most likely are oxidized in the
surface region. Third, the effective spin magnetic moment for Co derived from
the FY mode spectra shows the same magnetic field dependence as the macroscopic
magnetization measured by SQUID magnetometry. That is, the detailed magnetic
characterization of our Zn$_{0.95}$Co$_{0.05}$O films provides convincing
evidence that the room-temperature ferromagnetic-like behavior results from
superparamagnetic metallic Co clusters with diameters between 3 and 4\,nm. We
have no evidence for bulk room-temperature ferromagnetism resulting from
carrier mediated ferromagnetic exchange between diluted Co moments. The
interpretation of the nature of the ferromagnetic-like behavior in
Zn$_{0.95}$Co$_{0.05}$O in terms of superparamagnetic Co clusters is confirmed
by our detailed microstructural analysis. Both x-ray diffractometry and HRTEM
in combination with EFTEM directly prove the existence of metallic Co
nanoparticles with the same diameter as derived from the magnetic
characterization.

In summary, we identify metallic precipitates in Zn$_{0.95}$Co$_{0.05}$O thin
films as superparamagnetic cobalt clusters of nanometer size. We argue that the
magnetic behavior in our epitaxial cobalt-doped ZnO films is dominated by
these nanosized metallic Co clusters, leading to a ferromagnetic-like response
at room-temperature. The clusters are difficult to detect and can be revealed
only by a systematic element specific characterization. To this end, FY-XMCD
and thorough microstructural analysis are particularly valuable. We
emphasize that the formation and growth of the nanoparticles can be controlled
by the growth conditions and co-doping. In this way, it may be possible to
engineer the nanoparticles in a bottom-up technique and use them to tailor
material properties for specific applications as proposed recently for
Cr-rich ferromagnetic clusters in (Zn,Cr)Te \cite{Kuroda:2007a}.

\section*{Acknowledgements}

We thank Andreas Erb for the careful preparation of the polycrystalline
target materials for the pulsed laser deposition process.
This work was supported by the DFG via SPP~1157 (projects GR~1132/13 and
MA~1020/11), SPP~1285 (project GR~1132/14), and by the ESRF (project HE-2089).
Financial support of the German Excellence Initiative via the
\textit{Nanosystems Initiative Munich (NIM)} is gratefully acknowledged.

\clearpage

\begin{figure}[ht]
  \centering
  \includegraphics[width=10cm]{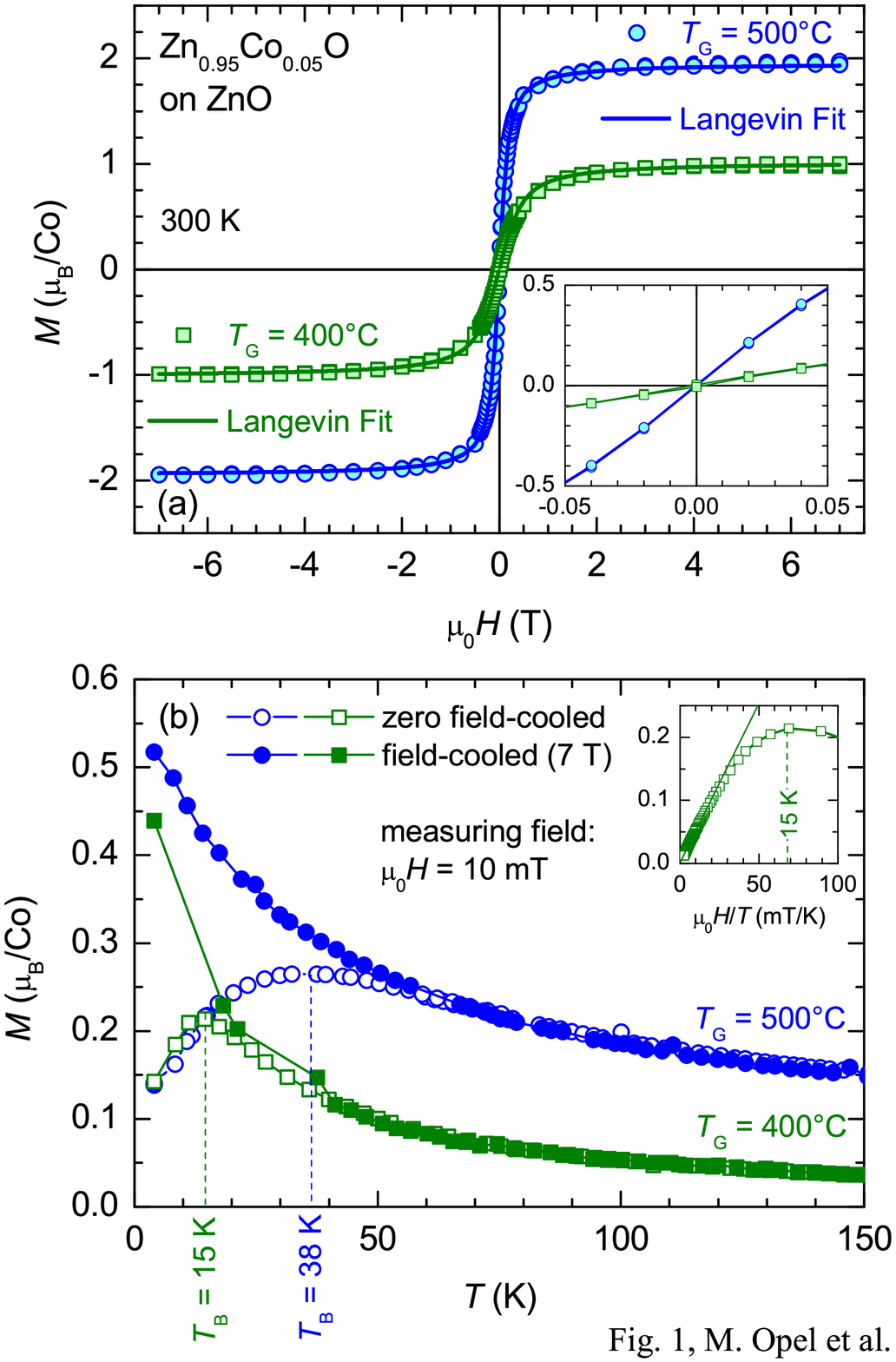}
  \caption{(color online) (a) Room temperature magnetization curves from
           Zn$_{0.95}$Co$_{0.05}$O thin films grown at
           $T_{\rm G} = 400^\circ$C (green squares) and $500^\circ$C (blue circles)
           with the magnetic field applied parallel to the film plane.
           The data can be fitted using the standard Langevin
           function of eq.~(\ref{eq:langevin}) with
           $\mu = 2370 \, \mu_{\rm B}$ (green line) and $5910 \, \mu_{\rm B}$ (blue line),
           respectively, indicating the presence
           of superparamagnetic particles in the samples.
           The inset shows the region around zero field on an enlarged scale.
           (b) Zero field-cooled (open symbols) and field-cooled
           magnetization measurements (closed symbols), taken
           at $\mu_0 H =$~10~mT as a function of temperature $T$. For both samples,
           the curves obtained after zero field-cooling show maxima at
           $T_{\rm B} = 15$~K and 38~K, respectively, pointing to a blocking
           of superparamagnetic particles at these temperatures.
           The inset shows the zero field-cooled magnetization vs $\mu_0 H/T$ for one sample
           for temperatures up to 375~K. The data nicely follow a Curie
           law (straight line) for temperatures well above $T_{\rm B}$.}
\end{figure}

\begin{figure}[ht]
  \centering
  \includegraphics[width=10cm]{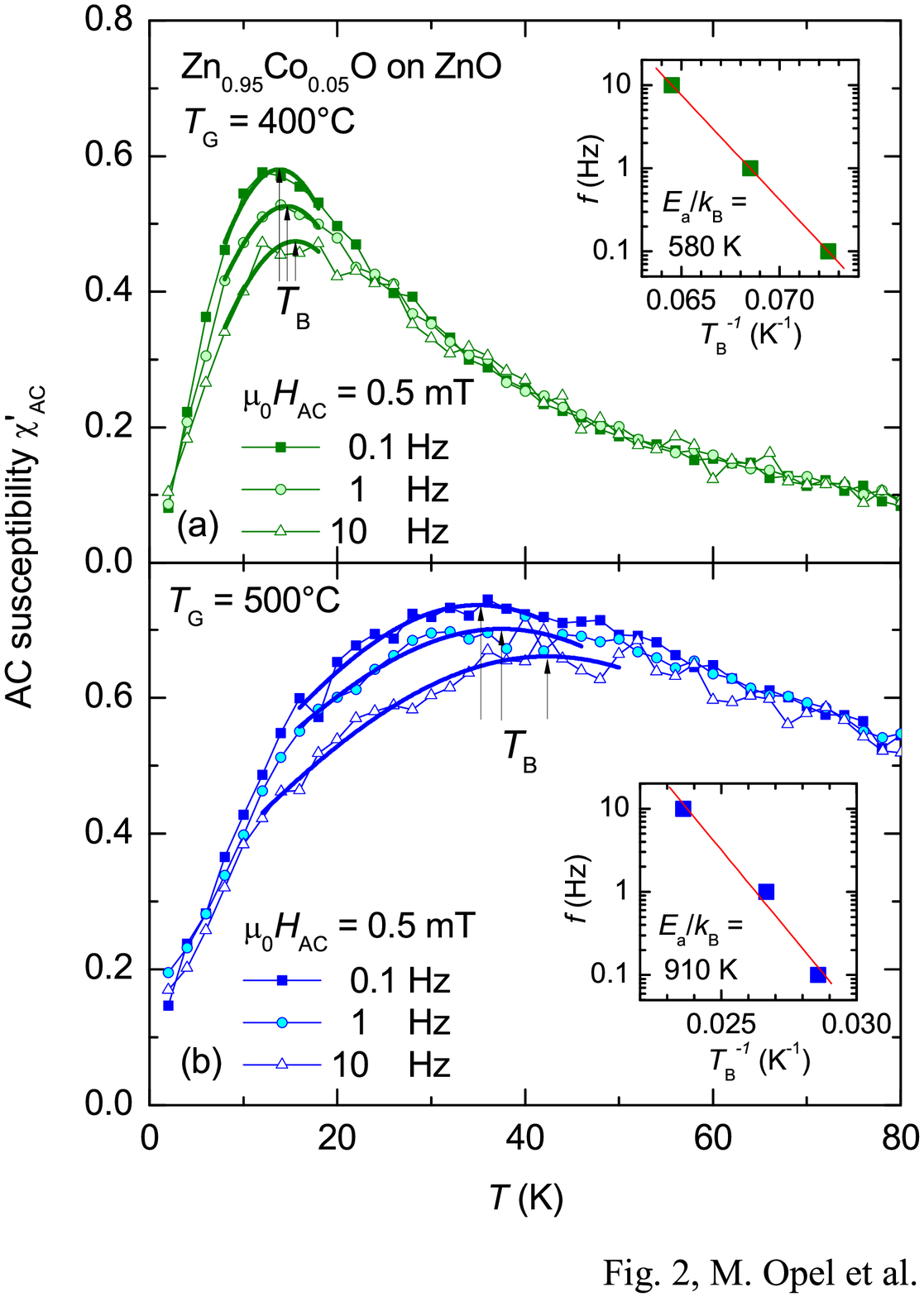}
  \caption{(color online) Real part $\chi'_{\rm AC}$ of the AC susceptibility (symbols)
           as a function of temperature $T$ from
           Zn$_{0.95}$Co$_{0.05}$O thin films grown at
           (a) $T_{\rm G} = 400^\circ$C (green) and (b) $500^\circ$C (blue).
           The lines are guides to the eye.
           The AC field of 0.5~mT was applied parallel to the
           film plane at frequencies $f = 0.1, 1, 10$~Hz. The positions
           of the maxima of the $\chi'_{\rm AC}(T)$ curves indicate the blocking
           temperature $T_{\rm B}$ (arrows).
           The insets show the frequency dependence of
           $T_{\rm B}$ (solid squares), which follows a N\'{e}el-Arrhenius law
           according to eq.~(\ref{eq:Neel-Arrhenius}) (straight lines).}
\end{figure}

\begin{figure}[ht]
  \centering
  \includegraphics[width=10cm]{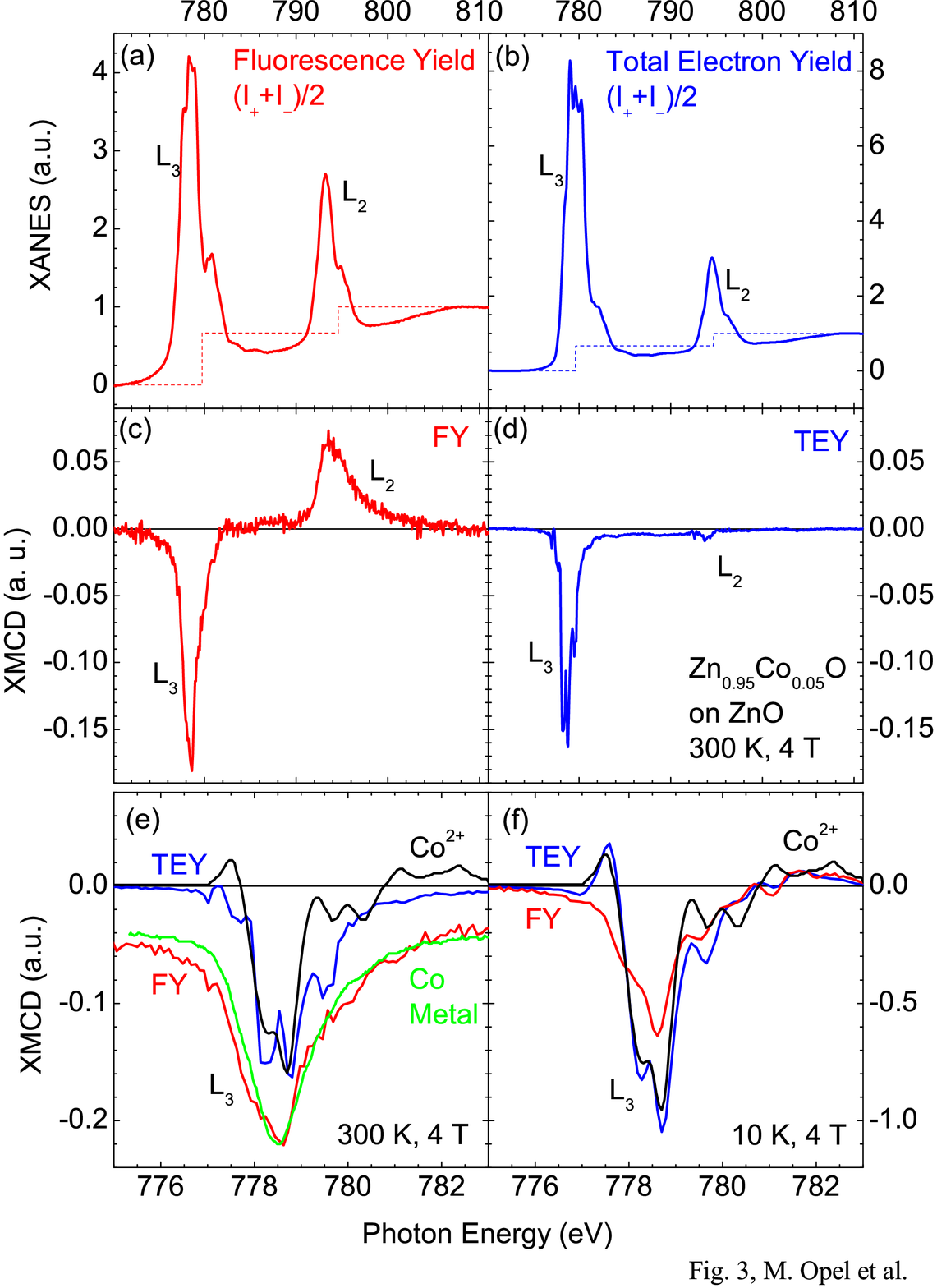}
  \caption{(color online) XANES spectra ((a) and (b)) and XMCD spectra ((c) and
           (d)) of a Zn$_{0.95}$Co$_{0.05}$O film grown at $T_{\rm G} =
           400^\circ$C. The data were measured in the fluorescence yield mode
           (FY, red, left panels) and total electron yield mode (TEY, blue, right panels),
           respectively, at 300\,K and an applied magnetic field of 4\,T.
           (e) and (f) show the region around the $L_3$ edge on an enlarged
           scale together with the XMCD calculated for Co$^{2+}$ (black, taken
           from \cite{Kobayashi2005}) and measured for metallic cobalt (green,
           \cite{Mamiya2006}). In (e), for clarity the curves for FY and Co
           metal have been shifted by $-0.04$\,a.u.}
\end{figure}

\begin{figure}[ht]
  \centering
  \includegraphics[width=10cm]{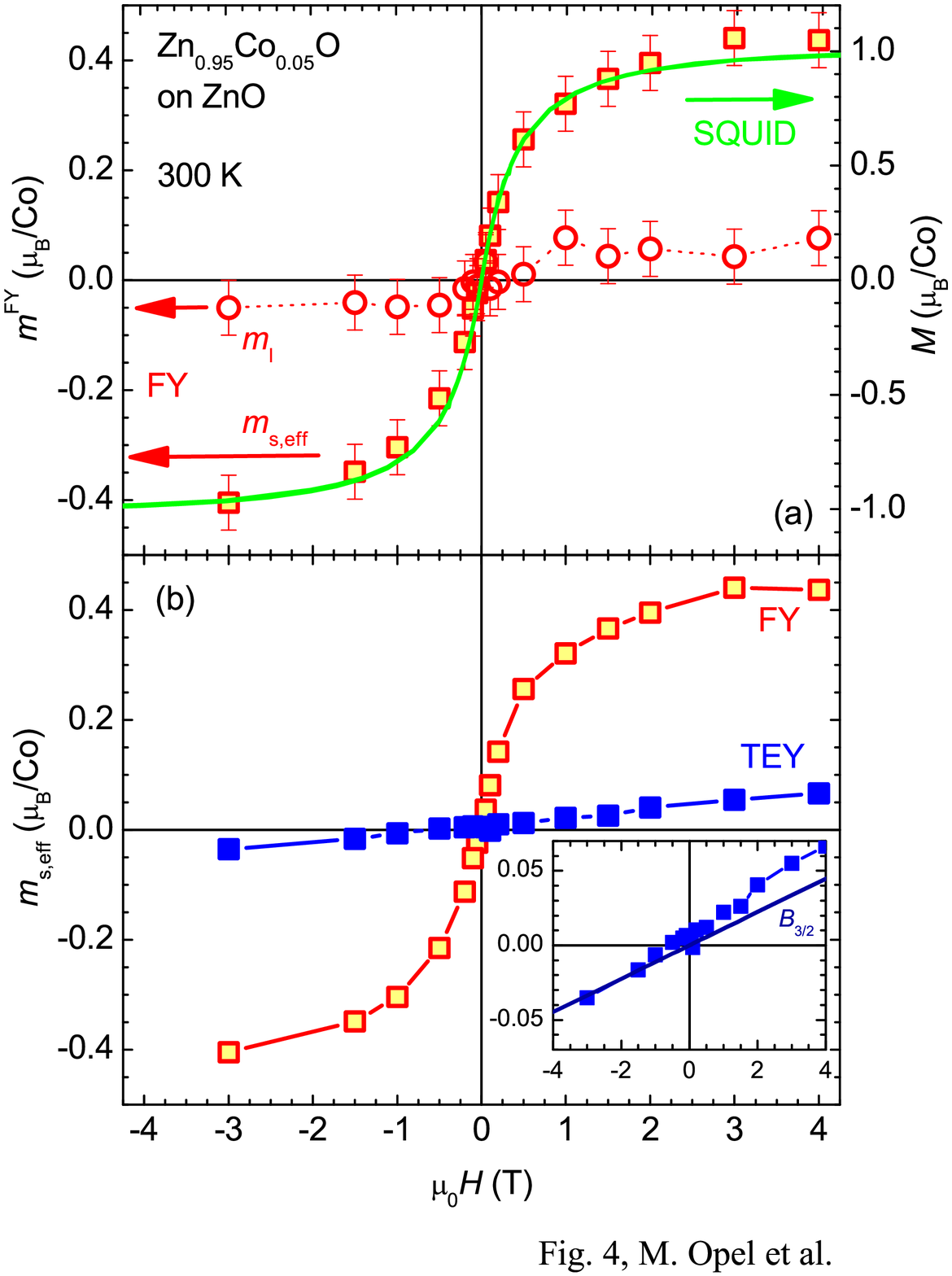}
  \caption{(color online) (a) Room-temperature effective spin magnetic moment $m^\mathrm{FY}_\mathrm{s,eff}$ (squares)
           and orbital moment $m^\mathrm{FY}_\mathrm{l}$ (circles) of Co in a
           Zn$_{0.95}$Co$_{0.05}$O thin film plotted versus the applied magnetic
           field. The moments are derived from the XMCD intensities (FY)
           using the magnetooptical sum rules. For comparison we also have plotted the magnetization
           $M(H)$ from Fig.1 measured by SQUID magnetometry
           (green line, right scale).
           (b) Effective spin magnetic moments $m_\mathrm{s,eff}$
           derived from the XMCD spectra recorded in the FY (red) and TEY
           (blue) mode at 300\,K. In the inset, $m^\mathrm{TEY}_\mathrm{s,eff}(H)$
           (blue) is compared to a Brillouin function (solid line) calculated for
           Co$^{2+}$ ions ($g=2$, $S=3/2$, $T=300$\,K).}
\end{figure}

\begin{figure}[ht]
  \centering
  \includegraphics[width=10cm]{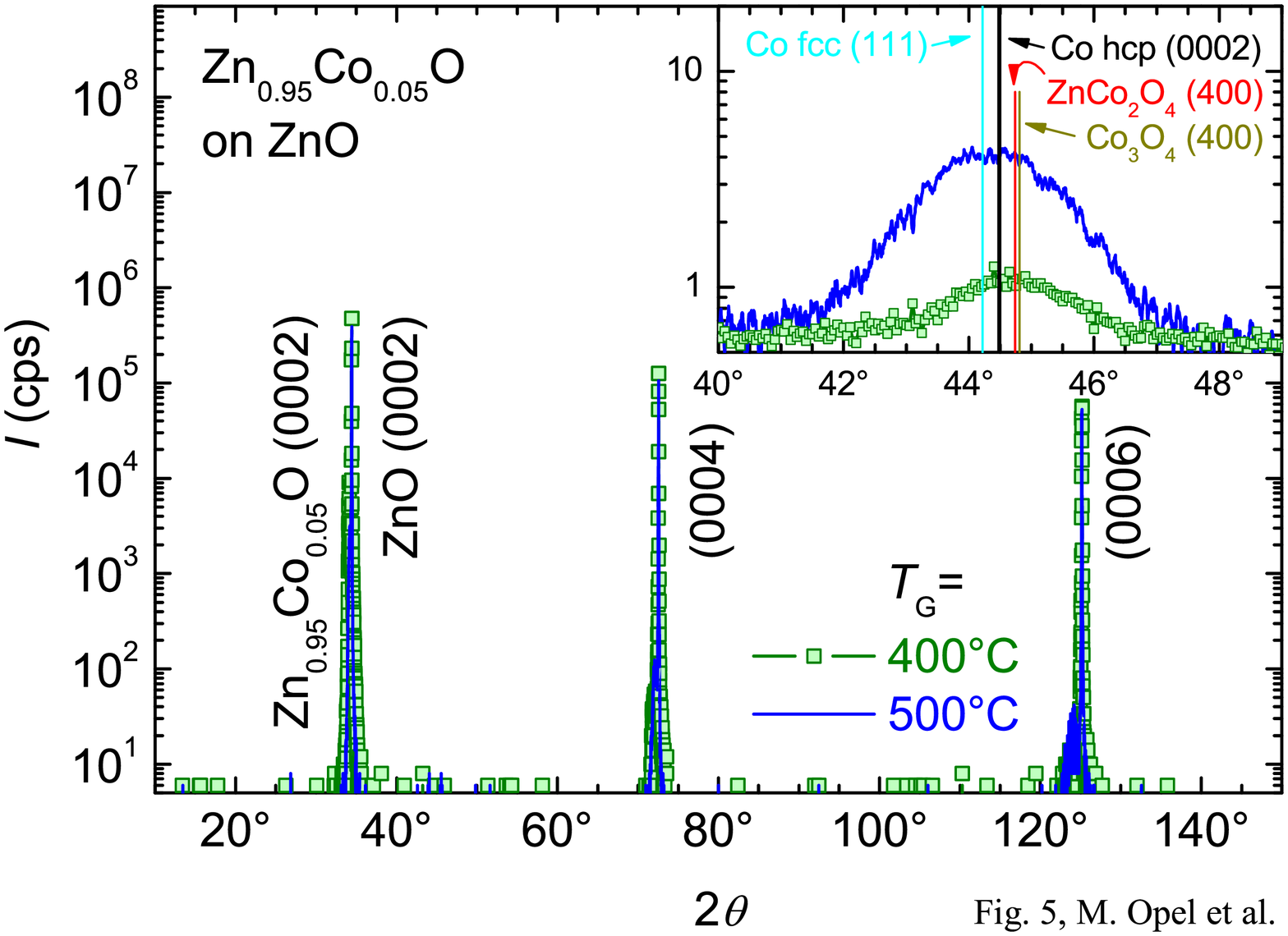}
  \caption{(color online) X-ray diffraction diagrams ($\omega$-$2\theta$ scans) from the Zn$_{0.95}$Co$_{0.05}$O thin
           films grown at $400^\circ$C and $500^\circ$C.
           The inset shows an enlargement of the region around $2\theta = 44^\circ$.
           After strongly increasing the integration time a minority phase reflection
           can be revealed at a position which agrees well with some peak positions expected for
           fcc or hcp metallic Co, ZnCo$_2$O$_4$ or Co$_3$O$_4$ (vertical lines).}
\end{figure}

\begin{figure}[ht]
  \centering
  \includegraphics[width=10cm]{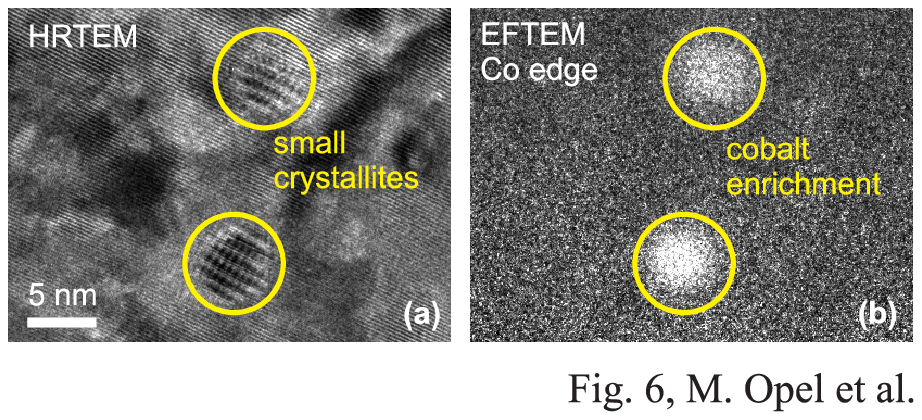}
  \caption{(color online) (a) Bright field TEM micrograph of a Zn$_{0.95}$Co$_{0.05}$O thin film
           in $[1\overline{1}00]$ orientation grown at 500$^\circ$C.
           Circles highlight regions with contrast originating from clusters
           with crystal structure different from ZnO.
           (b) Elemental map of Co clearly reveals Co enrichment at
           locations of the clusters.}
\end{figure}

\end{document}